\documentclass[aps,prb,showpacs,twocolumn,groupedaddress,amsmath,amssymb]{revtex4-1}
\usepackage{times,xspace}
\usepackage{amsfonts}
\usepackage{amsfonts}
\usepackage{bm}
\usepackage{amssymb}
\usepackage[final]{graphicx}
\usepackage{color}
\begin{document}
\title{Stripes, spin resonance and nodeless $d-$wave pairing symmetry in Fe$_2$Se$_2$-based layered superconductors}
\author{Tanmoy Das$^1$, and A. V. Balatsky$^{1,2}$}
\address{$^1$Theoretical Division, Los Alamos National Laboratory, Los Alamos, NM-87545, USA.}
\address{$^2$Center for Integrated Nanotechnology, Los Alamos National Laboratory, Los Alamos, NM 87545, USA}

\date{\today}
\begin{abstract}
We calculate RPA-BCS based spin resonance spectra of the newly discovered iron-selenide superconductor by using the two orbitals tight-binding model in 1 Fe unit cell. The slightly squarish electron pocket Fermi surfaces at $(\pi,0)/(0,\pi)-$momenta produce leading interpocket nesting instability at incommensurate vector $q\sim(\pi,0.5\pi)$ in the normal state static susceptibility, pinning a strong stripe-like spin-density wave or antiferromagnetic order at some critical value of $U$. The same nesting also induces $d_{x^2-y^2}-$pairing in one Fe unit cell. The superconducting gap is nodeless and isotropic on the Fermi surfaces as they lie concentric to the four-fold symmetric point of the $d-$wave gap maxima, in agreement with various experiments. This produces a slightly incommensurate spin resonance with upward dispersion, in close agreement with neutron data on chalcogenides. Finally, we demonstrate the conversion procedure from a 1 Fe unit cell to a 2 Fe one in which the gap symmetry transformed simultaneously into a $d_{xy}-$pairing and the resulting resonance spectra moves from the $q\simeq(\pi,\pi)$ to the $q\simeq(2\pi,0)/(2\pi,0)$ region.
\end{abstract}
 \maketitle\narrowtext

\section{Introduction}

The newly discovered high-$T_c$ superconductor in iron-selenide (Fe$_2$Se$_2$-) based compounds\cite{FeSe,FeSe1,FeSe2} not only increases the number in the unconventional superconducting (SC) family, but also bears on several unique magnetic and SC properties coming from its unusual Fermi surface (FS) topology. Both angle-resolved photoemission spectroscopy (ARPES)\cite{Zhang,HDing,Mou} and the local density approximation (LDA) calculations\cite{cao,cao1}, have demonstrated that  in Tl$_{0.63}$K$_{0.37}$Fe$_{2-x}$Se$_2$ at a critical value of Fe vacancy $x=0.22$ to tune the superconductivity, one can eliminate the hole-pocket at the $\Gamma$ point and the FS consists of only two electron pockets at $(\pi,0)/(0,\pi)$. Specific heat\cite{Cv}, and ARPES\cite{Zhang,HDing,Mou} studies find that these FSs host nodeless and isotropic SC gaps, whereas NMR measurements predict that the corresponding pairing is spin-singlet.\cite{WYu_NMR} Evidence is mounting from Raman,\cite{Raman} $\mu$SR,\cite{musr} optical,\cite{opticalSDW} NMR,\cite{NMR,NMR1} and other experiments\cite{yingsdw}, and supported by LDA calculations\cite{cao,cao1} that these materials possess a magnetic order ground state which coexists with superconductivity.

In cuprates, the FS is a hole-like pocket centering at the $k=(\pi,\pi)$ point and the nesting between inter FSs is strongest along $Q=(\pi,\pi)$ which is responsible for a spin-density wave (SDW), $d_{x^2-y^2}-$wave pairing, and a spin resonance peak at $Q$ with an `hour-glass' dispersion in the SC state.\cite{bourges,PLee} In a Ce-based heavy fermion, the hole pockets at $k=(\pi,\pi)$ lead to $Q=(\pi,\pi,\pi)-$resonance with $d_{x^2-y^2}-$wave symmetry.\cite{cecoin5} In iron-pnictide\cite{inosov} and iron chalcogenides,\cite{argyriouupward,Lihourglass,Lumsden45,QiuFeSeT} the FS has two hole-pockets at $\Gamma$ and two electron-pockets at $k=(\pi,0)/(0,\pi)$ in the unfolded 1 Fe unit cell and the leading nesting between the hole and the electron pockets constitute $q=(\pi,0)-$ SDW, $s^{\pm}-$ pairing, and $Q=(\pi,\pi)-$spin resonance with an `hour-glass' or upward dispersion (only observed in chalcogenides). In KFe$_2$As$_2$ only the hole pocket is present at $\Gamma$ and is predicted to have {\it nodal} $d-$wave pairing.\cite{KFeAs} On the contrary, in iron-selenide only one (or two concentric) electron-pocket(s) appeared at $k=(\pi,0)/(0,\pi)$ at a particular value of Fe vacancy.\cite{Zhang,HDing,Mou,cao,cao1} A similar FS topology appears in the particular case of electron-doped cuprate near underdoping where the antiferromagnetic (AFM) gap eliminates the hole-pocket in the nodal region, resulting in a nodeless $d$-wave gap.\cite{Das}

The role of FS topology and the interplay between electron and hole pockets and the exact shape of the neutron scattering intensity as a consequence of the change in the FS shape across these compounds remain a major challenge in the field. In this paper we investigate the unusually located electron pockets in iron-selenide superconductors in a two band tight-binding (TB) model and their role in  magnetic susceptibility. We find that the strong instability in the static susceptibility evolves around an incommensurate vector $q=(\pi,0.5\pi)$ at some critical value of $U$ leading to a stripe like SDW or AFM order in these systems, in agreement with experiments.\cite{NMR,NMR1,Raman,musr,opticalSDW,yingsdw} Unlike the stripe order in cuprates or other iron based superconductors, here the stripe SDW order has the same $q$ modulation at which a $d_{x^2-y^2}-$ pairing symmetry possesses a sign change at the hot-spots which results in a spin resonance peak at $Q\sim(\pi\pm\delta,\pi\pm\delta)$ in the SC state. Focusing mainly on Tl$_{0.63}$K$_{0.37}$Fe$_{1.78}$Se$_2$, and at the experimental SC gap value of 8.5~meV,\cite{HDing} we predict the resonance peak to be present in the range of 12.4meV at $Q\simeq0.78(\pi,\pi)$. The resonance profile also yields an `hour-glass' or upward dispersion and $45^o$ rotation of the resonance profile, in close agreement with observations in chalcogenides\cite{argyriouupward,inosov,Lihourglass} and cuprates.\cite{bourges}

An important aspect of identifying the $d-$wave pairing and the resonance is that all these results have to be converted simultaneously to the actual 2 Fe unit cell in which they will look as follows: (1) stripe order nesting $(\pi,0.5\pi)\rightarrow(0.5\pi,0.5\pi)$, (2) $d_{x^2-y^2}$-wave pairing$\rightarrow d_{xy}$-wave and (3) spin resonance at $(\pi\pm\delta,\pi\pm\delta)\rightarrow(2\pi\pm\delta,0)$. Furthermore, all these results are solely governed by the FS topology and thus they are reproducible with the inclusion of a higher number of bands away from the FS.

The paper is organized as follows. In Section II, we present our two-orbital TB formalism and the fitting of the dispersion and FSs to first-principle calculation and ARPES data, respectively. The evolution of the `stripe' competing order through the calculation of static susceptibility within the random phase approximation (RPA) calculations is given in Section III. The spin-resonance spectra  and the role of $d-$wave pairing are studied in Section IV. Section V gives a detailed discussion of how all the results of the 1 Fe unit cell Brillouin zone (BZ) transformed to the 2 Fe unit cell BZ. The temperature dependence of the spin-resonance spectra at $Q=(\pi,\pi)$ in 1 Fe unit cell is studied in Section VI. Using a five band model, we show in Section VII that all the results obtained with two-band model remains unchanged as the higher orbitals lie away from the Fermi level. Finally, we conclude in Section VIII.

%

\section{Tight-binding formalism}
First principle calculations show that the main contribution to the density of states (DOSs) near $E_F$ comes from $t_{2g}-$orbitals of Fe 3$d$ states which disperse only weakly in the $k_z-$direction.\cite{cao,cao1} Similarly to pnictide\cite{chimatrix,TB}, the role of the $d_{xy}$ orbit can be approximated by a next nearest neighbor hybridization between $d_{xz}$, $d_{zy}$ orbitals, and we consider a two-dimensional square lattice with two degenerate $d_{xz}$, $d_{zy}$ orbitals per site. Based on these observations, our model Hamiltonian\cite{TB} is
\begin{equation}\label{Hamiltonian}
H_0 = \sum_{{\bf k},\sigma}\psi^{\dag}_{{\bf k},\sigma}\left[\xi^+_{{\bf k}}{\bf 1} + \xi^-_{{\bf k}}{\bf \tau_3} + \xi^{xy}_{{\bf k}}{\bf \tau_1}\right]\psi_{{\bf k},\sigma},
\end{equation}
where $\sigma$ is the spin index, $\tau$ values are Pauli matrices and $\psi^{\dag}_{{\bf k},\sigma} = [d^{\dag}_{xz\sigma},d^{\dag}_{yz\sigma}]$ is the two component field operator. We consider up to third nearest-neighbor hopping for the present case as depicted in Fig.~1(b) which gives
\begin{eqnarray}\label{Hamiltonian}
\xi^{\pm}_{{\bf k}}&=&-(t_x^1\pm t_y^1)(c_x\pm c_y)-2(t_{xx}\pm t_{yy})c_xc_y\nonumber\\
&&~~-(t_x^2\pm t_y^2)(c_{2x}\pm c_{2y})-\mu^{\pm}\nonumber\\
\xi^{xy}_{{\bf k}}&=&-4t_{xy}s_xs_y.
\end{eqnarray}
Here $c/s_{i\alpha} = \cos/\sin{(ik_{\alpha}a)}$ for $\alpha=x,y$, $\mu^+=\mu$ and $\mu^-=0$. The Hamiltonian in Eq.~1 can be diagonalized straightforwardly obtaining the eigenstates $E_{0{\bf k}}^{\pm}=\xi^+_{\bf k}\pm\sqrt{(\xi^-_{{\bf k}})^2+(\xi^{xy}_{{\bf k}})^2}$.[Ref.~\onlinecite{TB}]
The TB parameters are obtained after fitting the dispersion to the LDA calculations,\cite{cao,cao1} [Fig.~1(c)]: $(t^1_x,~t^1_y,~t_{xx},~t_{yy},~t^2_x,~t^2_y,~t_{xy})=(-0.12,~-0.108,~-0.12,~-0.12,~0.048,~0.108,~0.06)$eV and $\mu=-0.36$eV. To obtain the experimental FS, we use a rigid band shift to $\mu=-0.56$eV and get two concentric squarish electron pockets centering $(\pi,0)$, in good agreement with ARPES\cite{HDing} [see Fig.~1(d)].

\begin{figure}[top]
\hspace{-0cm}
\rotatebox{0}{\scalebox{.72}{\includegraphics{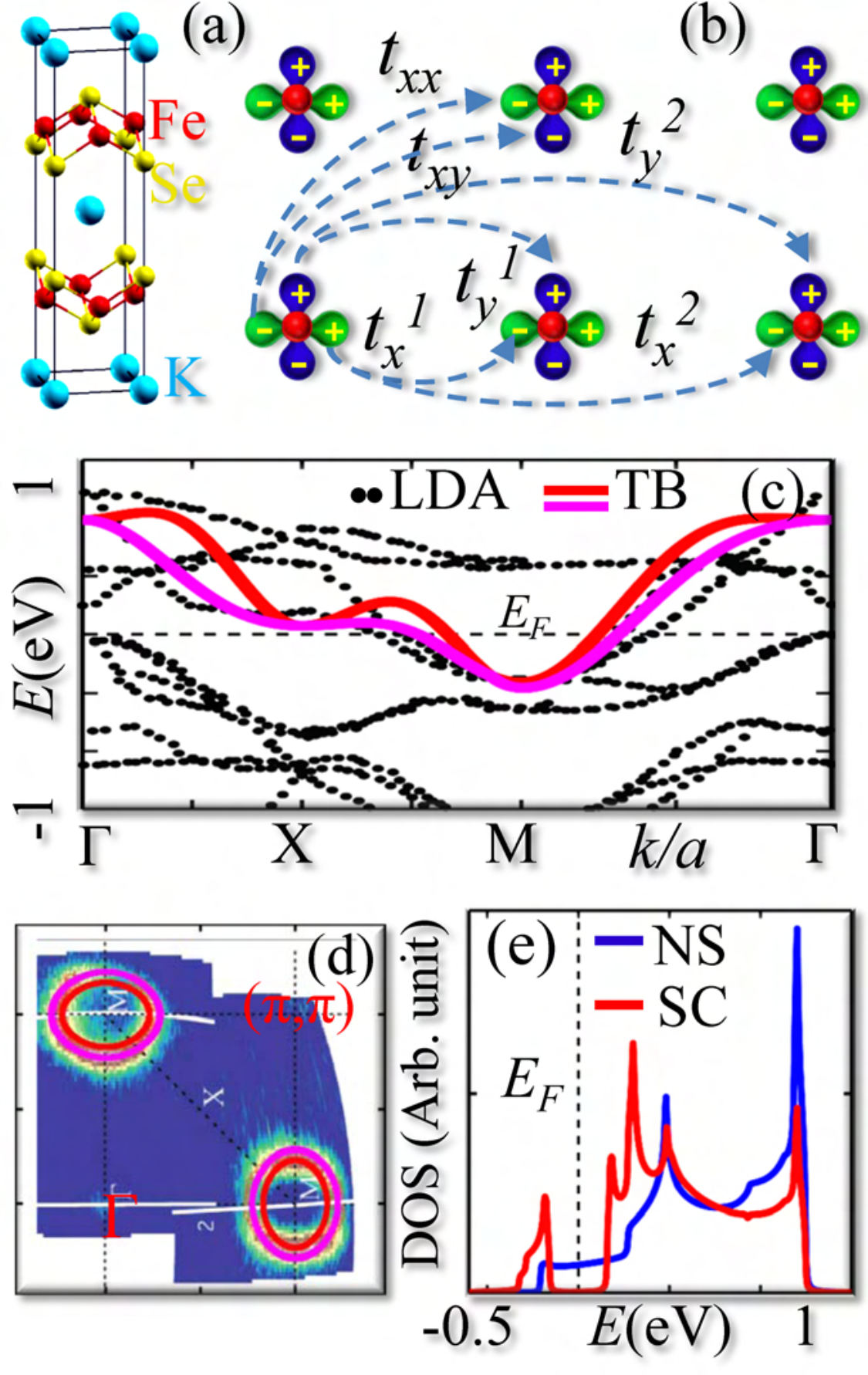}}}
\caption{(Color online) (a) Unit cell of a KFe$_2$Se$_2$-layered compound. (b) $x$-$y$ plane of the unfolded zone with Fe 3$d_{xz}$ and 3$d_{yz}$ orbitals. 
(c) The two resulting bands are plotted along high-symmetry momentum cuts and compared with the LDA one [dotted (black) lines]).\cite{cao,cao1} We mainly concentrate on fitting the low-energy part of the band to get the FS shape and their orbital characters in agreement with ARPES. (d) Computed electron pocket FSs overlain ARPES data\cite{HDing}. (e) The normal state (NS) DOS is compared with its SC counterpart. An artificially large value of gap=100meV is chosen here to highlight the nodeless and isotropic nature of the gap opening at $E_F$, despite $d-$wave pairing.} \label{fig1}
\end{figure}

As shown below, for such an FS, nesting along $Q=(\pi,\pi)$ is most pronounced leading to a $d_{x^2-y^2}-$pairing according to conventional views. We include an SC gap with the same amplitude on each band to obtain the SC quasiparticle dispersion as $E^{\pm}_{\bf k} = \sqrt{(E^{\pm}_{0{\bf k}})^2+\Delta^2_{\bf k}}$ where
%
$\Delta_{\bf k}=\Delta_0(\cos{k_xa}-\cos{k_ya})/2$.
$\Delta_0$ is taken from ARPES to be 8.5meV,\cite{HDing} for both eigenstates throughout this paper, except in Fig.~1(e) where an artificially large value of $\Delta_0$ is chosen to explicate the behavior of unconventional gap opening on the DOS. Despise the presence of nodes in the underlying gap symmetry, the dispersion at all momenta and the resulting DOS exhibit nodeless behavior in Fig.~1(e) as the FS pieces in these compounds are small in size and reside centering at the four-fold symmetric momenta of the gap maxima, similarly to pnictide but unlike in cuprate. The nodeless and isotropic gap is observed in ARPES\cite{Zhang,HDing,Mou} and specific heat measurements\cite{Cv}.

\begin{figure}[top]
\hspace{-0cm}
\rotatebox{0}{\scalebox{.52}{\includegraphics{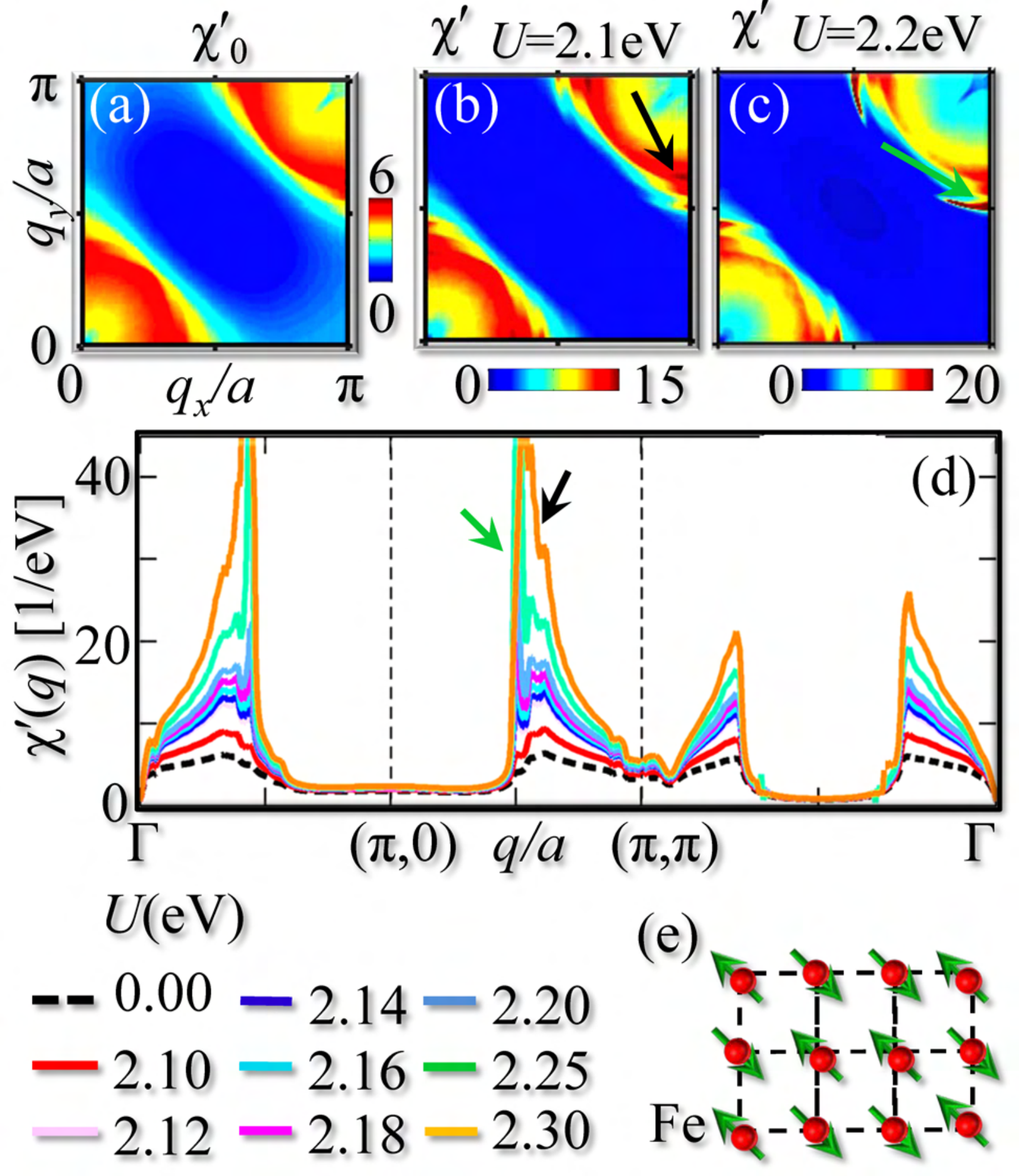}}}
\caption{(Color online) (a)
Bare static $\chi^{\prime}_{0}(q,\omega=0)$, and (b,c) its RPA part, $\chi^{\prime}(q)$, for two representative values of $U$. (d) $\chi^{\prime}(q)$ for different values of $U$ is drawn along the high-symmetry lines. The black and green arrows separate the peaks in $\chi^{\prime}(q)$ coming from the inner and outer FS pockets (see text). (e) Schematic top view of the FeSe layer with the Fe spins in the stripe AFM order expected in our calculation are shown by the arrows.} \label{fig1}
\end{figure}

\section{Static susceptibility and stripe ground state}

We now proceed to the spin-susceptibility, $\chi$, calculation of our model Hamiltonian. The calculation of the BCS $\chi$ for a multiband system including matrix-element effects of all the orbitals is standard and can be found for example in Refs.~\onlinecite{chimatrix,chiBCS,TB,Das2resonances}. We consider the electron-electron correlation on the same Fe atom within the RPA formalism.\cite{chimatrix} The terms in the RPA interaction vertex $U_s$ that are included in the present calculation are the intraorbital interaction $U$, an interorbital interaction $U^{\prime}=U-J/2$, the Hund's rule coupling $J=U/4$, and the pair hopping strength $J^{\prime}=J$.\cite{chiBCS}

Due to the specific nature of the FS shown in Fig.~1(d), the interpocket nesting is the dominant and incommensurate centering at $q=(\pi,\pi)$, see static bare $\chi^{\prime}_0(q)$ in Fig.~2(a). An additional nesting occurs centered at $q=0$ due to the intrapocket nesting of the semi-squarish portion of the FSs.
 While the shape of $\chi^{\prime}_0(q)$ is governed by the shape of the FS, the corresponding intensity is related to the Fermi velocity of the corresponding `hot-spots'.\cite{bob} Each of the $\chi^{\prime}_0(q)$ pattern is split into two due to the splitting of the electron pockets into two concentric ones within our model Hamiltonian.

Within the RPA, at different critical values of $U$, one or the other piece of $\chi^{\prime}(q)$ will dominate, leading to a SDW order phase, see Fig.~2(b)-2(e). To clarify that, we study the evolution of the RPA susceptibility at different values of $U$, keeping the other interactions same with respect to the value of $U$ (following the expressions given above). For a broad range of $U$ studied in this case, the incommensurate peak around $q=(\pi,\pi)$ is the winner, suggesting that a stripe like the SDW order will be present in this class of materials. These observations are consistent with the LDA calculations\cite{cao,cao1} as well as with number of experiments.\cite{NMR,NMR1,Raman,musr,opticalSDW,yingsdw} In the small range of $U$, the inner electron pocket gains more intensity (as in the case of $\chi_0$, indicated by black arrows), while the outer one is comparatively small (green arrows). Interestingly, when $U>2.2$eV, the situation is dramatically reversed and the outer nesting develops remarkable instability properties both around $q=(\pi,\pi)$ and $q=0$, although the former is always dominant in the range of $U$ studied here. With careful observation, one can find that the strong intensity within RPA is shifted toward $q=(\pi,0.5\pi)/(0.5\pi,\pi)$. This specific incommensurate vector is responsible for the possible $45^o$ rotation of the dynamical spin resonance spectra studied below. 

The small $q=(0,0.46\pi)/(0.46\pi,0)$ nesting which we find to be considerably large for our choice of parameters is close to the $q=0$ ferromagnetic (FM) instability or may induce some form of charge-density wave (CDW) with finite $q$ modulation. Note that the coexistence of a CDW and SDW due to multiple nestings is an emergent phenomenon predicted in pnictide\cite{CDWFeAs} and cuprates\cite{bob,Das2gap}. The presence of vacancy order in these systems can promote the coexistence of several competeting order.\cite{WBao2,DasVorder} Scanning tunneling microscopy (STM) may be able to detect it.

 Unlike the $(\pi,0)$-CDW order in cuprates or the $(\pi,0)$-SDW order in pnictide, here the $q=(\pi,0.5\pi)/(0.5\pi,\pi)$-stripe SDW order will facilitate $d$-wave pairing. This is due to the fact that the $q=(\pi,0.5\pi)/(0.5\pi,\pi)$-nesting comes from the inter-electron pockets at which the $d$-wave pairing possesses a change in sign of the gap, which is the criterion for spin resonance to occur as calculated below. But $q=(0,0.46\pi)/(0.46\pi,0)$ instability will provide pair-breaking contributions to the $d-$wave symmetry.

\begin{figure*}[top]
\hspace{-0cm}
\rotatebox{0}{\scalebox{.7}{\includegraphics{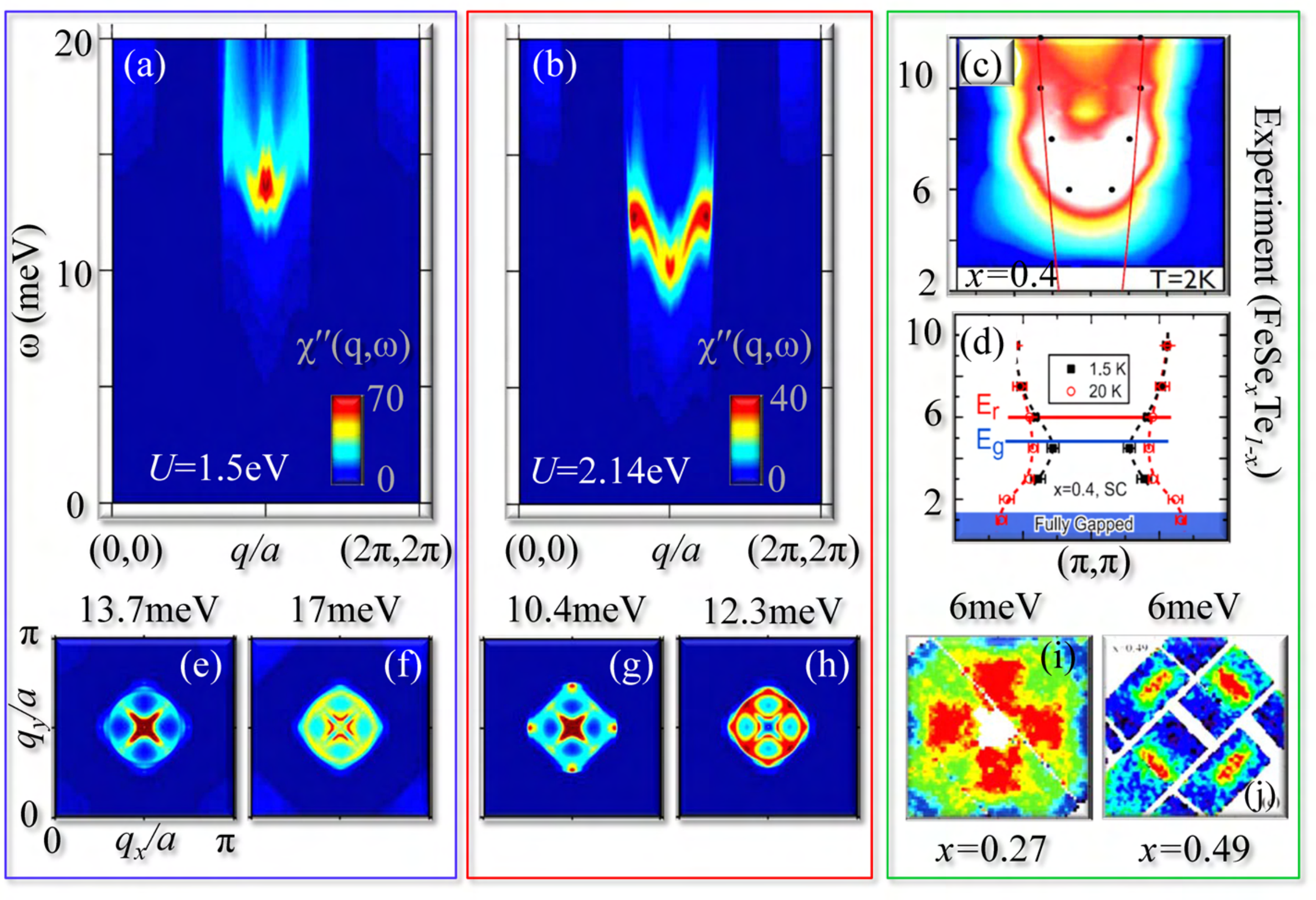}}}
\caption{(Color online) Computed BCS-RPA $\chi^{\prime\prime}(q,\omega)$ for (a) $U=1.5$eV and (b) $U=2.14$eV. The constant energy spectra at two representative energies are shown in the lower panels of the corresponding spectra. The right-hand column shows the neutron data on FeTe$_{1-x}$Se$_{x}$, demonstrating the observations of (c) an upward dispersion\cite{argyriouupward}, in (d) an `hour-glass' or `camel-back' pattern\cite{Lihourglass} and (i,j) $45^o$ rotation of the spectra\cite{Lumsden45} in this compound.} \label{fig1}
\end{figure*}

\section{Dynamical susceptibility, $d$-wave pairing and spin resonance}

Based on these findings, we proceed to study the evolution of the spin resonance spectra in the SC state, i.e., $\chi^{\prime\prime}({\bf q},\omega)$ (RPA-BCS), which is directly measured by an inelastic neutron scattering (INS) experiment. In the SC state, $\chi_0^{\prime}({\bf q},\omega)$ is strongly enhanced for $\omega<2\Delta$ due to turning on of the pair-scattering terms. This reduces the critical value of the interaction $U$ to satisfy the RPA denominator to be positive at $U\chi_0^{\prime}({\bf q},\omega)\le1$. We show the resonance spectrum for $U=1.5$eV in Fig.~3(a) and that for $U=2.14$eV in Fig.~3(b), which obtain a commensurate and an incommensurate resonance, respectively, in addition to their characteristic dispersion.

The dispersion of $\chi^{\prime\prime}({\bf q},\omega)$ which has a one-to-one correspondence with the dispersion of $\chi_0^{\prime}({\bf q},\omega)$, comes from three conditions\cite{norman,chiBCS,chimatrix,Das2resonances} in addition to the matrix-element effects: (1) $\omega_{res}({\bf q})=\sum_{k_F^{\nu},k_F^{\nu^{\prime}}}|\Delta_{{\bf k}_F^{\nu}}|+|\Delta_{{\bf k}_F^{\nu^{\prime}}}|$, (2) ${\rm sgn}(\Delta_{{\bf k}_F^{\nu}})\ne{\rm sgn}(\Delta_{{\bf k}_F^{\nu^{\prime}}})$, and (3) ${\bf q}={{\bf k}_F^{\nu}}+{{\bf k}_F^{\nu^{\prime}}}$ where $\nu,\nu^{\prime}$ are the band indices. Condition (1) comes from the energy conservation of inelastic scattering of Bogoliubov quasiparticles, while (2) constitutes the coherence factors of BCS susceptibility to possess a non-zero value. As both FS pockets have the same gap symmetries and amplitudes, and particularly for FeSe systems, where all the FS pockets have isotropic gaps at all ${\bf k}_F$ values, conditions (1) and (2) are satisfied equally at all $k_F$ values for inter electron pocket scatterings with $d_{x^2-y^2}-$symmetry. No intensity modulation in $\chi_0(q)$ is governed by conditions (1) and (2) and the only source of intensity variation is the interplay between the orbital matrix-element effect and condition (3), the latter is simply related to the number of degenerate ${\bf k}_F$ values. This means that the upward dispersion of the resonance behavior in FeSe can not be understood by tracking how the gap varies on the FS as happenes in cuprates,\cite{norman} but only from the FS nesting and the matrix-element in the tensor form of $\chi_0$, and, also, on the interaction vertex $U_s$.\cite{chimatrix}

\begin{figure*}[top]
\hspace{-0cm}
\rotatebox{0}{\scalebox{.8}{\includegraphics{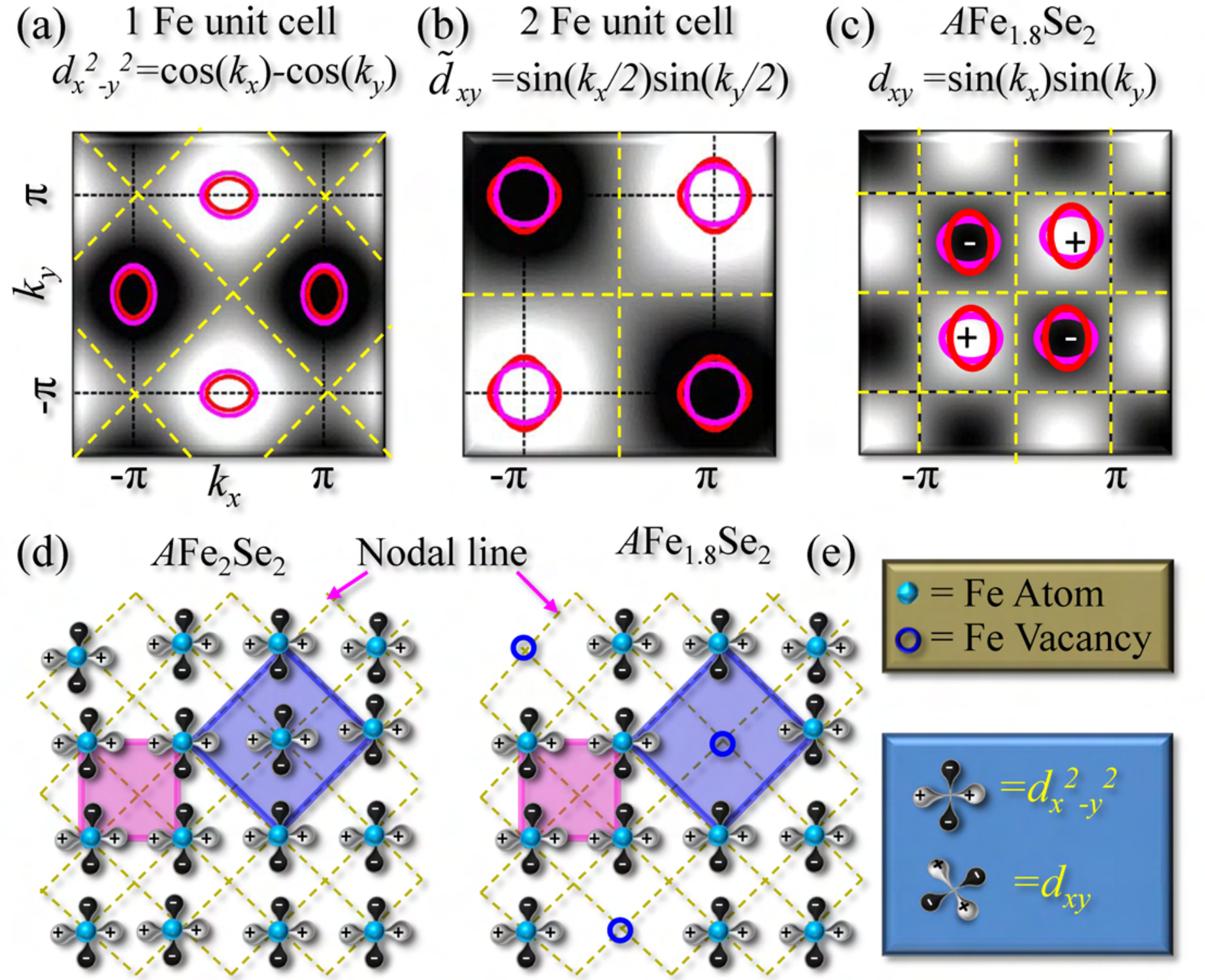}}}
\caption{(Color online) Correspondence between (a) the 1 Fe unit cell and (b) 2 Fe unit cell in the FS topology and SC gap symmetry. The black to white color scale denotes the negative to positive value of the gap. (c) Transformation of the $d_{xy}$-wave gap and the FS are plotted in the same 2 Fe unit cell as in (b) but with one Fe vacancy. (d,e)  Real space view of the gap-symmetry corresponding to momentum space ones is given in (b) and (c). The $d_{x^2-y^2}$-wave in the 1 Fe unit cell (magenta box) automatically takes the form of a ${\tilde d}_{xy}=\sin{(k_xa/2)}\sin{(k_ya/2)}$-wave in the 2 Fe unit cell (blue box). (e) Vacancy order obtained from the experimental data in Ref.~\onlinecite{WBao}. Yellow dashed lines in each plot denote the nodal lines for the corresponding pairing symmetry. } \label{fig4}
\end{figure*}

Furthermore, unlike in cuprates, where the `hour-glass' dispersion extends to $\omega=0$ due to the presence of nodal quasiparticles on the FS, in the nodeless FS of FeSe-compounds the resonance behavior spans only within $\Delta$ to $2\Delta$. For $U=1.5$eV, the maximum intensity lies at the bottom of the spectrum at the commensurate vector at $\omega_{res}=13.5$meV with $\omega_{res}/2\Delta=0.79$. Again, $U=2.14$eV, the maximum intensity shifts to the top of the spectra at an incommensurate vector $q=0.78(\pi,\pi)$ at $\omega_{res}=12.4$meV with $\omega_{res}/2\Delta=0.73$, slightly larger than its average value of $0.64$ found in all other superconductors.\cite{greven} The upward dispersion of the magnetic spectra is qualitatively analogous to the one observed by Neutron in FeTe$_{0.6}$Se$_{0.4}$ superconductors\cite{argyriouupward} shown in Fig.~3(c), while pnictides show more commensurate resonance within the experimental resolution.\cite{inosov} By including the weak-intensity of the dispersion below the $(\pi,\pi)$- resonance, one can draw an `hour-glass' or `camel-back' feature, also seen in chalcogenides\cite{Lihourglass} in Fig.~3(d).

The constant energy profile of the resonance spectra is more interesting and bears important physical insights. At $\omega\ge\omega_{res}$ for both values of $U$, the resonance profile is squarish with maximum intensity lying along the diagonal and finite intensity along the (100) direction, see Figs.~3(f) and 3(h). For $U=1.5$meV, the intensity remains concentrated at $(\pi,\pi)$ even when $\omega\le\omega_{res}$. Interestingly, for $U=2.14$eV at $\omega\le\omega_{res}$ the peak along the diagonal sharply looses intensity while the intensity peak along the bond direction remains almost the same. As a result, the resonance profile is rotated by 45$^o$ as one goes down below the resonance. The $45^o$ rotation of the resonance spectra is observed earlier in hole doped cuprates\cite{bourges} and can also be seen in chalcogenides between two different dopings in the SC state\cite{Lumsden45} in Fig.~3(i) and 3(j).

\section{One Fe unit cell to two Fe unit cell conversion}

It has been argued that when FS pockets are translated from a 1 Fe unfolded zone to a 2 Fe folded zone, it switches location from $k=(\pi,0)/(0,\pi)$ to $k=(\pi,\pi)$. Thus in a $d_{x^2-y^2}-$symmetry, the SC gap on the FS is assumed to change from nodeless to nodal behavior.\cite{Mazin} We show that this apparent inconsistency is merely a result of misinterpretation coming from not performing the same unitary transformation in the gap symmetry as in the FS. To perform this transformation, we denote a quantity with a tilde in the 2 Fe unit cell.

The unitary transformation consists of 45$^o$ rotation of the lattice with $\tilde{a}=\sqrt{2}a$ which gives $\tilde{k}_{x/y}=(k_x\pm k_y)/2$. In doing so, we find that the pairing symmetry $d_{x^2-y^2}=\cos{(k_xa)}-\cos{(k_ya)}$ in the 1 Fe unit cell transforms to an extended ${\tilde d}_{xy}=\sin{(k_xa/2)}\sin{(k_ya/2)}$, see Figs.~4(a) and 4(b).  This transformation can be more clearly understood from the real space view of the gap symmetry in the lattice in Fig.~4(d). The 1 Fe unit cell and 2 Fe unit cell are highlighted by magenta and blue boxes, respectively. It is clear that the nodal line passes through the diagonal direction in the former case, which automatically becomes aligned along the zone boundary in the latter unit cell. The rotation of the SC gap phases follows similarly. We note that such unitary transformation preserves the nodeless and isotropic properties of the SC gap in any unit cell.

An important consequence of the ${\tilde d}_{xy}$-wave SC gap is that it breaks the crystal symmetry. As can be seen from the momentum space view of the pairing phases in Fig.~4(b), for example, as we translate from $k = (-\pi/a,-\pi/a)$ to $k=(\pi/a,-\pi/a)$ by the reciprocal vector $2\pi/a$, the phase of the SC gap is reversed. One can recover the translational symmetry of the crystal by moving to $k = (2\pi/a,-\pi/a)$ by a reciprocal vector of $4\pi/a$. In the real space approach, this would correspond to a case when one Fe atom from the center of the 2 Fe unit cell is removed; see Fig.~4(e). Remarkably, such a crystal symmetry can be achieved when the vacancy order and the magnetic order conspire with each other to favor the $d_{xy}$-wave SC gap. This vacancy arrangement was found indeed to occur in the real system.\cite{WBao2} In this context, we note that superconductivity occurs in this class of materials when a suitable amount of vacancy is achieved which forms order and the crystal symmetry hops from the $I4/mmm$ to the $I4/mm$ group.\cite{WBao,DasVorder}

\begin{figure}[top]
\hspace{-0cm}
\rotatebox{0}{\scalebox{.6}{\includegraphics{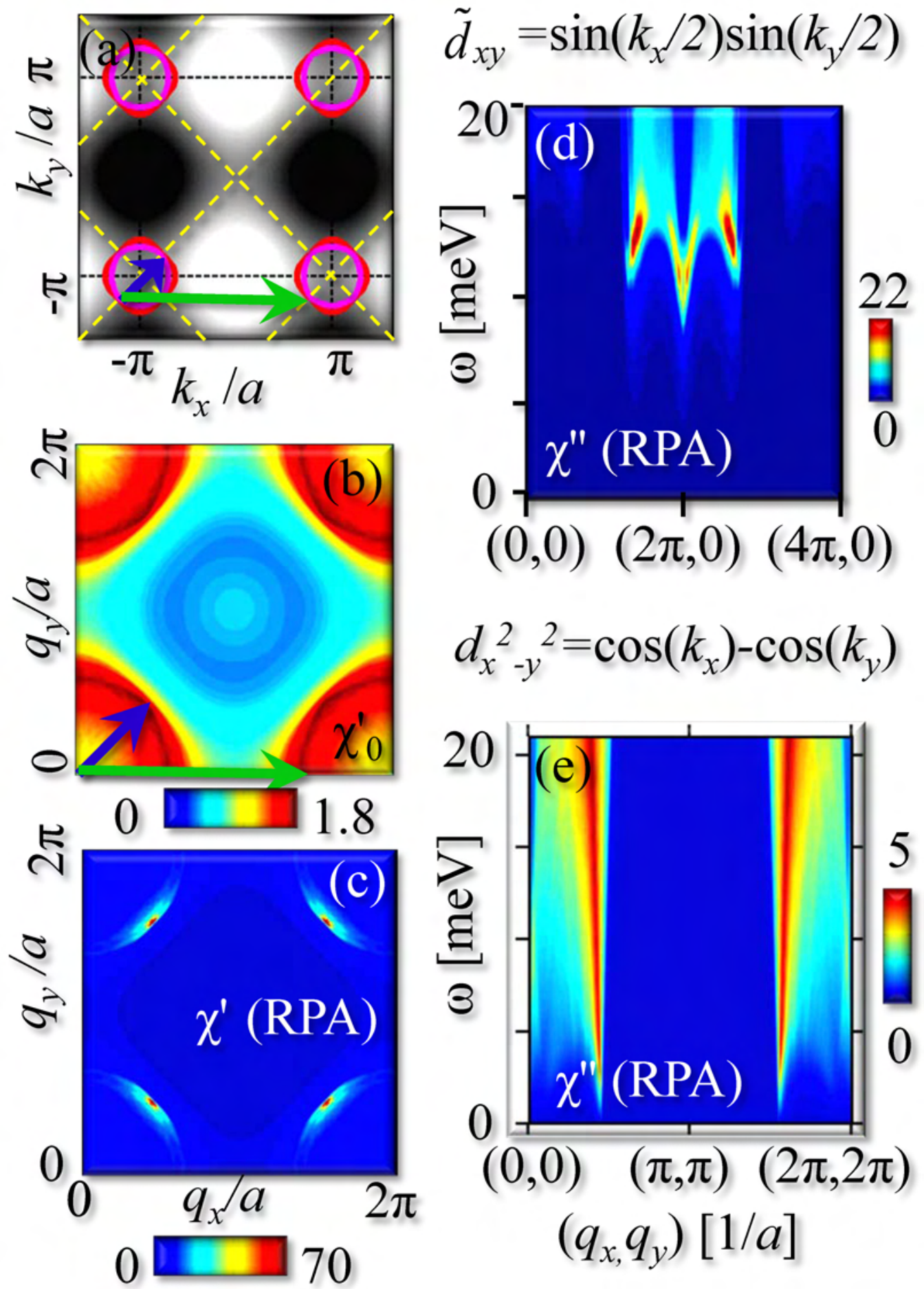}}}
\caption{(Color online) (a) FSs in the 2 Fe unit cell BZ are shown with various nesting channels (arrows). (c) Bare static susceptibility, $\chi_0^{\prime}$, is plotted in a 2 Fe unit cell. (c) Corresponding RPA instability at $U=1.7$eV. The unitary transformation leads to both nesting vectors of the 1 Fe unit cell, shown in Fig.~2(a), coinciding around $q\sim(\pi/2,\pi/2)$. (e) The dynamical spin suscebtibility, $\chi^{\prime\prime}$ is calculated at $U=1.8$eV as shown along the zone boundary [green arrow in (b)] for $\tilde{d}_{xy}$ pairing in the 2 Fe unit cell [$d_{x^2-y^2}$ in the 1 Fe unit cell]. (e) Computed spin-susceptibility is plotted for nodal $d_{x^2-y^2}$-pairing in 2 Fe unit cell [$d_{xy}$ in 2 Fe unit cell]. Here $U$ is chosen to be 1.5~eV.} \label{fig1}
\end{figure}

\subsection{Rotation of spin resonance in 2Fe unit cell}


The same unitary transformation leads to both the nesting vectors of 1 Fe unit cell shown in Fig.~2(a) to coincide with each other around $q=(\pi/2,\pi/2)$ in the 2 Fe unit cell as shown by blue arrow Figs.~5(a)-5(b). Subsequently the nesting along the (200)-direction shown by green arrow serves as the `hot-spot' to the ${\tilde d}_{xy}$, where the SC gap changes signs. This leads to resonance spectra around $(2\pi\pm\delta,0)/(0,2\pi\pm\delta)$ as shown in Fig.~5(d).

Note that the smaller intrapocket `hot-spot' vector (blue arrow) will give a sign change of the SC gap on the same FS pocket and thus a nodal SC gap [$d_{x^2-y^2}$ in the 2 Fe/unit cell and $d_{xy}$ in the 1 Fe unit cell] will occur. The resulting INS spectrum in a nodal SC state is confined near ${\bf q}\sim0$ and does not show any true resonance peak but a spin-excitation dispersion as shown in Fig.~5(e). In this case the associated weak magnetic excitation strength may not be sufficient to produce large values of $T_c$ in these materials.

\begin{figure}[top]
\hspace{-0cm}
\rotatebox{0}{\scalebox{.5}{\includegraphics{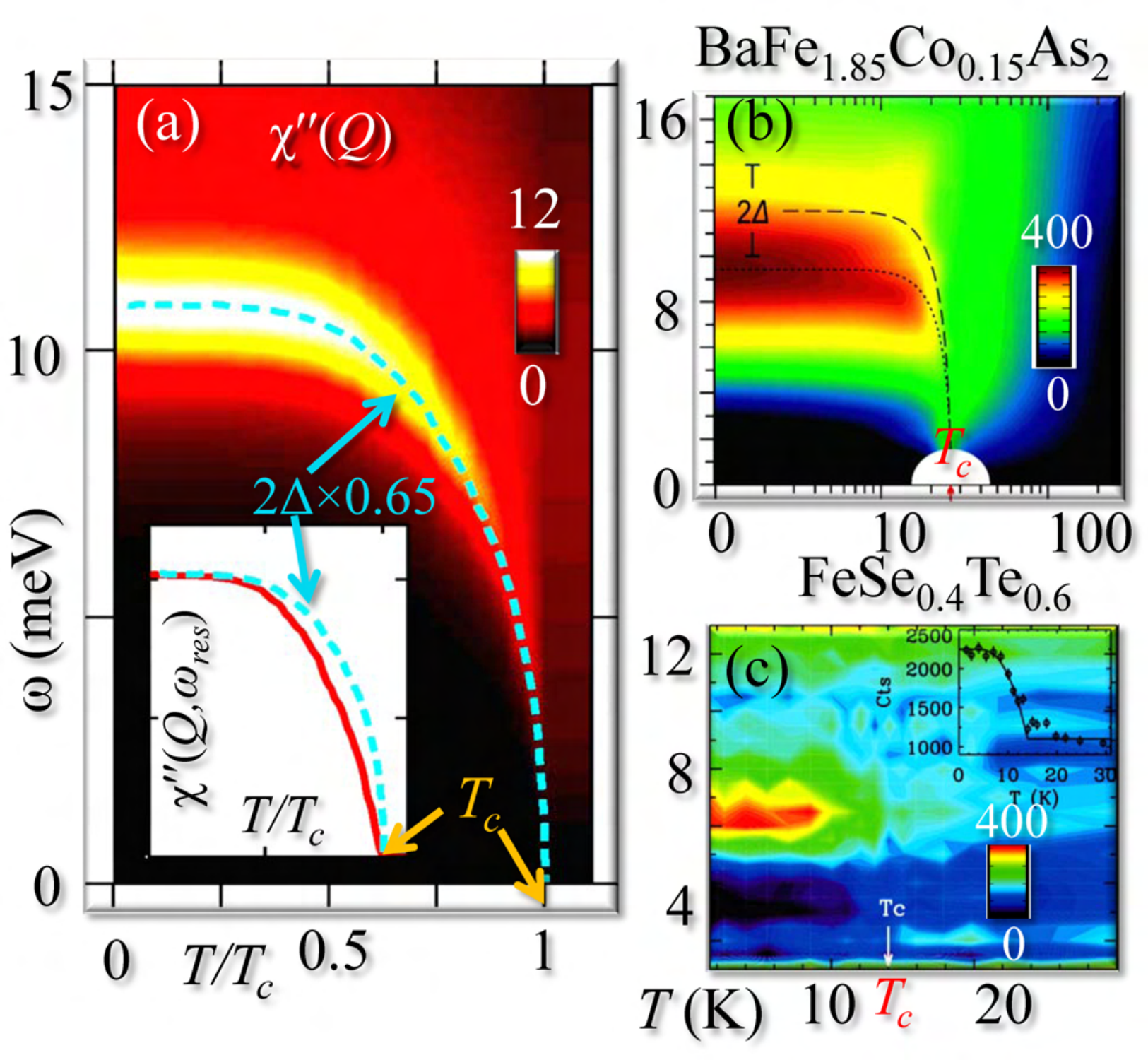}}}
\caption{(Color online) (a) Temperature dependence of the spin excitation spectrum at $q=(\pi,\pi)$ for $U=1.5$eV is compared with the similar observations by inelastic neutron scattering measurements (b) in pnictide\cite{inosov} and (c) chalcogeniges\cite{QiuFeSeT}. Inset: Computed value of the intensity of the resonance peak is compared with BCS $\Delta(T)$.} \label{fig1}
\end{figure}

\section{Temperature dependence of the spin-resonance at $Q$}

 In pnictide and chalcogenide superconductors  the temperature evolution of the spin resonance follows the BCS mean-field behavior of the SC gap $\Delta(T)$.\cite{QiuFeSeT,inosov}. We predict that similar phenomenon also occurs in FeSe-superconductors. To prove that, we calculate the $\Delta(T)$ for the two band system solving the standard BCS gap equation with phenomenological pairing strength parameters $V_{1,2}$= 52, 46~meV (neglecting interband pairing). The parameters are adjusted to obtain an experimental gap of $\Delta_0$=8.5meV with BCS ration $2\Delta/k_BT_c$=6.7, in close agreement with the experimental value of 7.\cite{HDing} $\Delta(T)$ is shown in Fig~6(a) (cyan line) on top of the calculated RPA-BCS spin resonance spectrum at $q=(\pi,\pi)$ with $U=1.5$eV. The intensity at the resonance is plotted in {\it inset} to Fig.~6(a) in red line. Both the resonance energy and the intensity shows a remarkable one-to-one correspondence to the value of $\Delta(T)$. The results agree well with those for pnictide and chalcogenide shown in Figs.~6(b) and 6(c) respectively. The $T-$dependence of the resonance energy and its intensity follows the BCS form of $\Delta(T)$, suggesting that the itinerant RPA-BCS Fermi-liquid theory will be valid in FeSe-based superconductors in particular and in all iron based superconductors studied to date in general.

\begin{figure}[top]
\hspace{-0cm}
\rotatebox{0}{\scalebox{.5}{\includegraphics{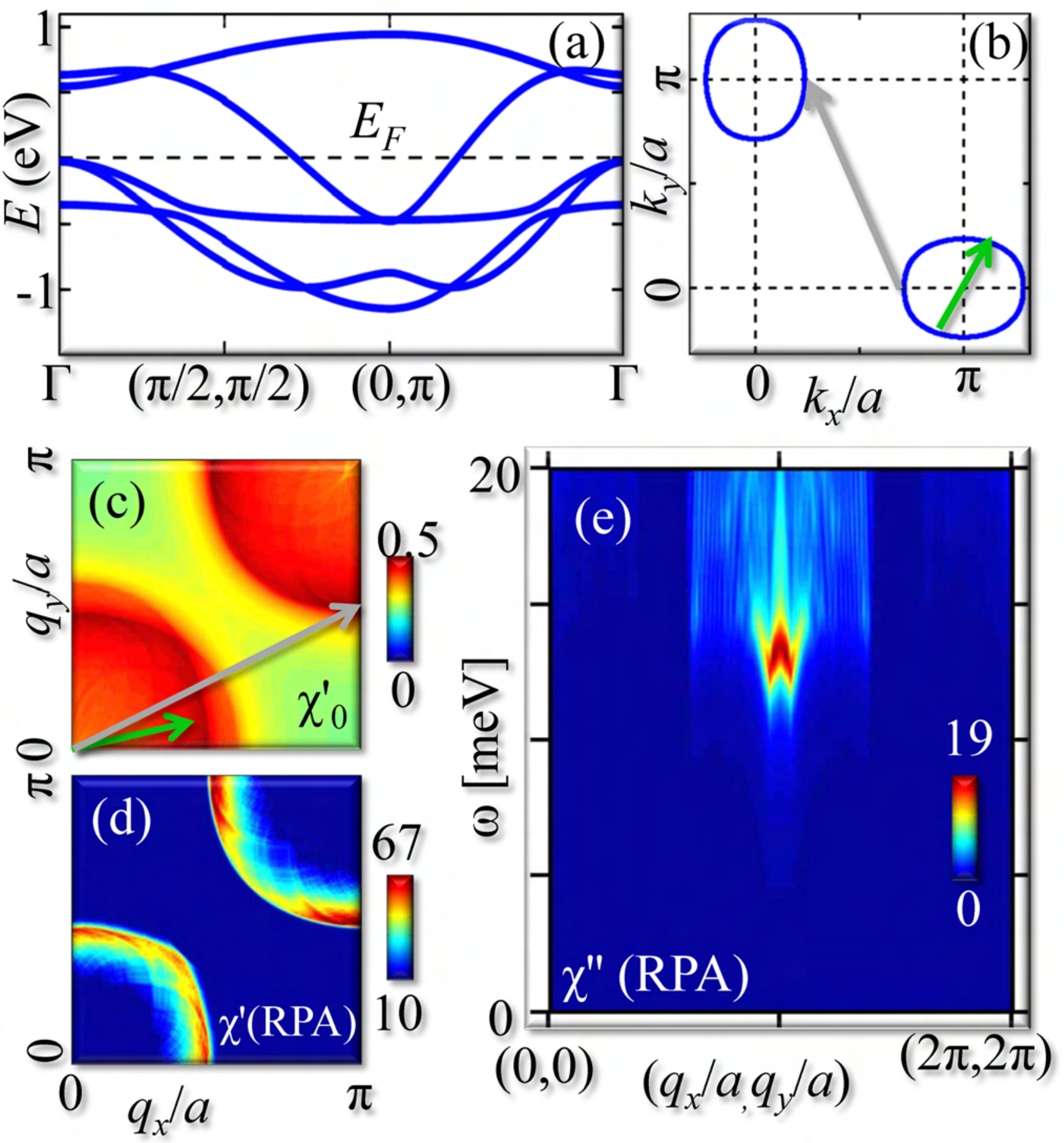}}}
\caption{(Color online) Calculations with the five-band model in the 1 Fe unit cell.\cite{wang}  (a) Calculated band-structure is drawn along the high-symmetry lines for a chemical potential $\mu=7.45$eV. (b) The corresponding electron-pocket FS is plotted in the first quadrant of the Brillouin zone. (c)-(d) The bare and the RPA component of the static spin susceptibility, respectively. (e) Computed spin excitation spectrum is shown along the diagonal direction in a $d_{x^2-y^2}-$wave pairing symmetry. These results should be compared with the results obtained with the two band model in the 1 Fe unit cell in Figs.~1-3. The value of $U$ for all RPA calculations here is set to be 1.5~eV.} \label{fig1}
\end{figure}

\section{Five band model}
We repeat the calculation presented in Figs.~1-3 within a five-band model and find that the magnetic structure and magnetic resonance do not depend on the overall band structure but are essentially governed by the FS topology. The TB model for 5 bands is taken from Ref.~\onlinecite{wang} which includes the 5 $d$ orbitals of the Fe atoms in the 1 Fe unit cell. The band structure is given in Fig.~7(a) and the corresponding FS is shown in Fig.~7(b). The topology of the FS is qualitatively captured here, although the five band model does not incorporate the nearly degenerate FSs as calculated in the LDA. Nevertheless, this minor technical drawback does not alter our conclusions.

The magnetic structure including the two nesting vectors at $q=(\pi,0.5\pi)/(0.5\pi,\pi)$ and $q=(0,0.46\pi)/(0.46\pi,0)$ is well reproduced in the five band calculation as shown in Figs.~7(b) and 7(c). Furthermore, the magnetic resonance spectra is remarkably close to the one obtained in Fig.~3. The subtle differences between the two band and the five band calculation come from the inconsistency in the FS topology, not in the overall band dispersion. The remarkable consistency between the two band and the five band model suggests that the magnetic ground state as well as the superconducting properties can be studied properly once all the FS pieces are modeled accurately.

\section{Conclusion}

In summary, we have calculated the static and dynamical spin susceptibility for an effective two orbital TB model with $d$-superconductivity. We find that the leading instability in $\chi^{\prime}$ is evolved around an incommensurate vector $q=(\pi,0.5\pi)$ at some critical value of $U$, pinning stripe like SDW or AFM order, in agreement with a number of experiments.\cite{NMR,NMR1,Raman,musr,opticalSDW,yingsdw} Unlike in other iron based superconductors, here the SDW order has the same $q-$vector for the $d$-pairing symmetry to obtain the opposite sign at the `hot-spot'. Our choice of parameters also predict a comparatively weak CDW modulation with $q=(0,0.46\pi)$, similar to cuprates,\cite{bob,Das2gap} and pnictides.\cite{CDWFeAs} The $d-$wave gap yields a spin resonance behavior with an `hour-glass' or upward dispersion and the $45^o$ rotation of the resonance profile, in close agreement with observations in chalcogenides. The present results also demonstrate that the weak to moderate pair coupling theory of the spin fluctuation mechanism of superconductivity is also appropriate in FeSe-based superconductors in particular and in all iron based superconductors studied to date in general. A similar intermediate coupling theory has been successful in describing many salient features in cuprates.\cite{Dasop,DasNFL}

Recently, we became aware of the works by Wang {\it et al.}\cite{wang} and Maier {\it et al.}\cite{maier} which also predicted the $d-$wave gap in FeSe based systems.

\begin{acknowledgments}
We are grateful to H. Ding, A. V. Chubukov, and W. Bao for useful discussion. This work was funded by US DOE, BES and LDRD and benefited from the allocation of supercomputer time at NERSC.
\end{acknowledgments}

\end{document}